\documentstyle[preprint,aps]{revtex}
\begin{document}
\preprint{MKPH-T-01-22, TRI-PP-01-34}
\draft
\tighten
\title{The anomalous chiral perturbation theory meson Lagrangian to order 
$p^6$ revisited}
\author{T.\ Ebertsh\"auser,$^1$ H.~W.~Fearing,$^2$ S.~Scherer$^1$}
\address{$^1$ Institut f\"ur Kernphysik,
Johannes Gutenberg-Universit\"at, J.~J.~Becher-Weg 45,
D-55099 Mainz, Germany\\[1ex]
$^2$ TRIUMF, 4004 Wesbrook Mall, Vancouver, British Columbia,
Canada V6T 2A3}
\date{18 October 2001}
\maketitle
\begin{abstract}
   We present a revised and extended construction of the mesonic Lagrangian 
density in chiral perturbation theory (ChPT) at order $p^6$ in the anomalous 
(or epsilon) sector, ${\cal{L}}_{6,\epsilon}$. 
   After improving several aspects of the strategy we used originally, i.e., 
a more efficient  application of partial integration, the implementation of 
so-called Bianchi identities, and additional trace relations, we find the new 
monomial sets to include 24 $SU(N_f)$, 23 $SU(3)$, and 5 $SU(2)$ elements. 
   Furthermore, we introduce 8 supplementary terms due to the extension of 
the chiral group to $SU(N_f)_L\times SU(N_f)_R\times U(1)_V$.
\end{abstract}

\section{Introduction}
   Chiral perturbation theory for mesons 
\cite{Weinberg_79,Gasser_84,Gasser_85} has proven to be a highly successful 
method for describing the interactions of the pseudoscalar octet 
($\pi,K,\eta$) at low energies (for an overview of recent activities see, 
e.g., Ref. \cite{Bernstein_98}). 
   It is based on a chiral $SU(3)_L\times SU(3)_R$ symmetry of the QCD 
Lagrangian in the limit of massless $u$, $d$, and $s$ quarks in combination 
with the assumption of a spontaneous symmetry breaking to 
$SU(3)_V$ in the groundstate.
   According to Goldstone's theorem 
\cite{Nambu_60,Goldstone_61,Goldstone_62}, 
each broken generator, i.e., each generator which does not annihilate the 
vacuum, gives rise to a massless Goldstone boson whose properties are tightly 
connected to the generator in question.
   In the present case, one expects eight pseudoscalar Goldstone bosons 
transforming as an octet under $SU(3)_V$ with vanishing interactions in 
the limit of zero energies.
   These Goldstone bosons are identified with the low-energy pseudoscalar 
octet, where the finite masses of the physical multiplet result from an 
explicit symmetry breaking due to the  finite $u$, $d$, and $s$ quark masses 
in the underlying QCD Lagrangian. 
   As an analogous situation holds for $SU(2)$, we first consider the general 
case of $SU(N_f)$ and specialize to $N_f=2$ and $3$ later. 

   The effective Lagrangian density of ChPT is organized as a string of terms 
with an increasing number of covariant derivatives and quark mass terms,
\begin{equation}
\label{leff}
{\cal L}={\cal L}_2 + {\cal L}_4 + {\cal L}_6 +\cdots,
\end{equation}
where the subscripts refer to the order in the momentum and quark mass 
expansion. 
   The derivative expansion essentially takes account of the vanishing 
interaction at low energies with the explicit symmetry breaking being treated 
perturbatively. 
   At each order the most general Lagrangian compatible with chiral symmetry, 
parity, and charge conjugation invariance is required. 
   Here, we work in the framework of ordinary ChPT, where the quark mass 
term is counted as ${\cal O}(p^2)$, or, in other words, matrix elements are 
treated at a fixed ratio 
$m_{\mbox{\footnotesize quark}}/p^2$ \cite{Gasser_84}.
   For an overview of a different organization procedure we refer the reader 
to Ref.\ \cite{Stern_98}. 
   A systematic treatment of physical matrix elements is made possible by 
Weinberg's power counting scheme \cite{Weinberg_79}. 
   For example, at ${\cal O}(p^4)$ one has to consider tree-level diagrams 
with exactly one vertex from ${\cal L}_4$ and an arbitrary number of vertices 
from ${\cal L}_2$ or one-loop diagrams with vertices from ${\cal L}_2$. 
   The most general structure of ${\cal L}_4$ was first discussed by Gasser 
and Leutwyler \cite{Gasser_85} and contains 10 low-energy coupling constants 
$L_i$. 
   Out of these, 8 are required to absorb infinities generated by one-loop 
diagrams from ${\cal L}_2$. 
   The finite pieces are not predicted by chiral symmetry and have to be 
determined from experimental data.

   The ChPT action functional, which is (apart from the special case of the 
Wess-Zumino-Witten term) the four-dimensional space-time integral over some 
chirally invariant Lagrangian density, generates at any given chiral order a 
finite-dimensional real vector space. 
   The finite dimension is due to the fact that the basic building blocks 
can only be multiplied together in a finite number of different ways. 
   Unfortunately, there seems to be neither a way to predict this dimension 
beforehand nor a general algorithm to decide whether a set of given 
structures is linearly independent or not.
   That is the reason why it is almost impossible to tell if a generating 
system is linearly independent so that it actually represents a basis of 
the above mentioned vector space.
   To our knowledge even the ${\cal L}_4$ of \cite{Gasser_85} has not 
formally been shown to be a basis, though it is without any doubt a 
generating system and countless calculations seem to confirm that the 
terms are independent.
   In Ref.~\cite{Fearing_96} we systematically wrote down the most general 
Lagrangian density of chiral order ${\cal{O}}(p^6)$, both for the normal 
($\rightarrow$ 111 $SU(3)$ terms) and anomalous 
($\rightarrow$ 32 $SU(3)$ terms) sector. 
   Although these sets are also most likely to be generating systems,
the normal sector was later shown to include redundant structures 
\cite{Bijnens_99}.

   In the present paper, we provide a revised and slightly modified list of 
$SU(N_f)$ terms for the anomalous or epsilon sector, as well as its reduction 
to $SU(2)$ and $SU(3)$. 
   For the sake of clarity, we repeat the analysis from the start,
using a different notation which is closer to what has become the standard
in this field. 
   The initial ingredients and the general strategy are the same as used 
before.
   The main improvements in comparison to our former work are outlined one 
after the other in the following sections.
   In Sec.\ II, after defining different basic building blocks, 
which lead to nice simplifications of the partial-integration method 
(see Sec.\ III), we repeat the main features of our strategy.
   The so-called Bianchi identities for field strength tensors are introduced 
and implemented in Sec.\ IV.
   Section V contains a somewhat different approach to trace relations.
   In Sec.\ VI, additional structures triggered by the extension of the 
chiral group  to $SU(N_f)_L\times SU(N_f)_R\times U(1)_V$ are worked out.
  A concluding comparison of the anomalous Lagrangian density to chiral order 
$p^6$ derived in Ref.~\cite{Fearing_96} with the present work appears in 
Sec.\ VII.
   Section VIII gives a brief summary. Tables and eliminating relations 
are relegated to the Appendices.

\section{Modified basic building blocks and strategy}
   In Ref.~\cite{Fearing_96} a systematic construction of chiral Lagrangian 
densities was developed starting from structures of the type
\begin{eqnarray}
[A]_\pm\equiv\frac{1}{2}(A U^\dagger\pm U A^\dagger),
\end{eqnarray}
where $A$ is supposed to transform under the chiral group $G$ as 
$A\stackrel{G}{\rightarrow} V_R A V_L^\dagger$.
   The partial-integration argument to be outlined in Sec.~III may be 
simplified considerably by using a modified kind of basic building blocks 
\begin{eqnarray}
(A)_\pm\equiv 2u^\dagger [A]_\pm u,
\end{eqnarray}
which has already been used in a slightly different form by several 
authors (see, e.g., \cite{Bijnens_99,Akhoury_91}).
   Here, $u$ is defined via $u^2\equiv U$.
   Note that $(A)_\pm$ obeys the transformation rule 
$(A)_\pm\stackrel{G}{\rightarrow} V (A)_\pm V^\dagger$, 
where $V\equiv\sqrt{V_R U V_L^\dagger}^{-1}V_R\sqrt{U}$ is the so-called 
compensator.
   The covariant derivative 
\begin{eqnarray}
\label{covder}
\nabla_\mu(A)_\pm\equiv\partial_\mu(A)_\pm+[\Gamma_\mu,(A)_\pm]\quad
(\mbox{N.B.} : 
\nabla_\mu(A)_\pm\stackrel{G}{\rightarrow} V \nabla_\mu(A)_\pm V^\dagger)
\end{eqnarray}
induced by the chiral connection 
$\Gamma_\mu=\frac{1}{2}\left[u^\dagger,\partial_\mu u\right]
-\frac{i}{2}u^\dagger R_\mu u-\frac{i}{2}u L_\mu u^\dagger$ 
provides a key element for an easier and more efficient application of the 
total-derivative procedure, which is to be outlined in more detail shortly.

   Since we always have to take single or multiple traces (denoted by
$\langle \cdots \rangle$) to obtain 
Lagrangian density monomials, our final results will only differ by a number 
compared to the former structures
\begin{eqnarray}
\langle(A_1)_\pm\dots(A_m)_\pm\rangle
=2^m\langle[A_1]_\pm\dots[A_m]_\pm\rangle .
\end{eqnarray}
   Let us illustrate this fact by looking at the leading-order non-anomalous 
Lagrangian density
\begin{eqnarray}
{\cal{L}}_2=-\frac{F_0^2}{16}\langle 
{(D_\mu U)}_-{(D^\mu U)}_-\rangle+\frac{F_0^2}{4}\langle{(\chi)}_+\rangle
=-\frac{F_0^2}{4}\langle {[D_\mu U]}_-{[D^\mu U]}_-\rangle
+\frac{F_0^2}{2}\langle{[\chi]}_+\rangle.\nonumber
\end{eqnarray}

   As pointed out in Ref.~\cite{Fearing_96}, it is sufficient to restrict 
oneself to $[D^m U]_-$, $[D^n G]_+$, $[D^n H]_+$, 
and $[D^n \chi]_\pm$ ($m, n$ integer with $m>0, n\ge 0$). 
   So, we simply have to substitute for those ingredients the new building 
blocks $(D^m U)_-$, $(D^n G)_+$, $(D^n H)_+$, and $(D^n \chi)_\pm$.
   Here we use the definitions 
$G^{\mu\nu}\equiv F^{\mu\nu}_RU +U F^{\mu\nu}_L$ and 
$H^{\mu\nu}\equiv F^{\mu\nu}_RU-U F^{\mu\nu}_L$, 
where $F^{\mu\nu}_{L/R}$ are evidently the field strength tensors belonging 
to the corresponding external fields $R_\mu$ and $L_\mu$. 
   The $\chi\equiv 2B_0(s+ip)$ block contains the external scalar and 
pseudoscalar fields, and $B_0$ is related to the scalar quark condensate 
$\langle 0|\bar{q}q|0\rangle$ in the chiral limit.
   To simplify the comparison with other groups 
\cite{Fearing_96,Bijnens_99,Akhoury_91},
we have given a translation prescription in Table \ref{tp}.

   The so-called hierarchy strategy of Ref.\ \cite{Fearing_96} has proven to 
be extremely fruitful. 
   Collecting $(D^n G_{\mu\nu})_+$, $(D^n H_{\mu\nu})_+$, and 
$(D^n \chi)_\pm$ for a moment in one symbol $(D^n \chi_{\mu\nu})_\pm$, 
one immediately finds that all possible terms at 
${\cal{O}}{(p^6)}$ can either include no, one, two, or three $
(D^n \chi_{\mu\nu})_\pm$ blocks which naturally defines four distinct 
sectors or levels to be considered. 
   We say: the more $(D^n \chi_{\mu\nu})_\pm$ blocks are included in one 
monomial, the lower its level.
   Once all terms one can think of (at each level) are listed, one tries to 
generate relations to eliminate as many structures as possible. 
   We always try to get rid of terms as high in the hierarchy as 
possible.
   In particular, with this strategy one ensures that the number of terms is
minimal for the special case in which all external fields are set equal to zero.\footnote{The non-anomalous final $SU(N_f)$ set given in 
Ref.~\cite{Bijnens_99} is not minimal when setting all external fields to 
zero. In that case the structures $Y_1$ and $Y_4$ are not independent and can 
be eliminated. However, if we replace these two monomials by the terms (120) 
and (139) of Ref.~\cite{Fearing_96}, then the entire set will remain 
independent whether or not there are external fields and both structures, 
(120) and (139), vanish explicitly when the fields are set to zero.}
  
   Before explaining the improvements mentioned above in more detail, 
let us discuss the symmetry of multiple covariant derivatives. 
   It turns out to be more useful not to assume the multiple covariant 
derivatives to be symmetrized from the very beginning. 
   We implement the relations among unsymmetrized structures noting that 
just one general (i.e., non-contracted) index combination is actually 
independent in the sense of the hierarchy explained above.
   For double derivatives this statement reads
\begin{eqnarray}
(D_\mu D_\nu A)_\pm=(D_\nu D_\mu A)_\pm+\frac{i}{4} 
[ (A)_\pm , (G_{\mu\nu})_+ ] - \frac{i}{4} \{ (A)_\mp , (H_{\mu\nu})_+ \}.
\end{eqnarray}
   Although there would not be any disadvantage in keeping these 
unsymmetrized monomials in our final set 
(N.B. : $(D_\mu D_\nu U)_-$ will be the only one to be kept), 
for aesthetical reasons we replace them by symmetrized ones making use of 
the formula
\begin{eqnarray}
(D_\mu D_\nu U)^s_-\equiv\frac{1}{2}(\{D_\mu, D_\nu\} U)_-
=(D_\mu D_\nu U)_-  + \frac{i}{2} (H_{\mu\nu})_+ .
\end{eqnarray}

   We are now ready to initiate the construction procedure.
   First of all, we write down all conceivable 
(we find 74, see Table~\ref{structure_list}) anomalous structures 
(satisfying $P$ and $C$ invariance, hermiticity, and chiral order $p^6$) 
in terms of our chirally invariant basic building blocks.
   We then collect as many relations as possible which follow from any of 
the mechanisms we are aware of. 
   Those are 
\begin{enumerate}
\item[(1)] partial integration, 
\item[(2)] equation-of-motion argument, 
\item[(3)] epsilon relations,
\item[(4)] Bianchi identities,
\item[(5)] trace relations.
\end{enumerate} 

   As detailed descriptions of methods (2) and (3) can be found 
in Ref.~\cite{Fearing_96}, we only comment on the new or improved 
items (1), (4), and (5).
   The resulting final set is shown in Table \ref{final_set}. 
   There, the structures are again ordered according to the organization 
scheme introduced in IV.B of \cite{Fearing_96}. 
   All explicit non-redundant relations are listed in Appendices A to E. 
   There is no unique way to choose which of a set of monomials to keep. 
   We basically follow the example of \cite{Fearing_96}, i.e., in addition to 
the hierarchy approach we always try to get rid of multiple derivative terms, 
especially of the type $(D^2 U)_-$. 
As emphasized above, and in Ref.~\cite{Fearing_96}, since one cannot prove 
the independence of the final set of monomials, there is no guarantee that 
another approach might not happen to find additional relations leading to a 
smaller final number. Yet we have not found an alternative way, giving rise 
to further reductions.

\section{Partial integration  (total-derivative argument)}
   Recalling the fact that a total derivative in the Lagrangian density does 
not change the equation of motion, we can generate relations of the following
type
\begin{eqnarray}
& &\partial_\mu\langle(A_1)_\pm\dots (A_m)_\pm \rangle
+\overbrace{\langle[\Gamma_\mu,(A_1)_\pm\dots (A_m)_\pm ]\rangle}^{0}
=\langle\nabla_\mu 
[ (A_1)_\pm\dots (A_m)_\pm ]\rangle\nonumber\\
&=&\langle\nabla_\mu (A_1)_\pm\dots (A_m)_\pm\rangle+\dots
+\langle(A_1)_\pm\dots \nabla_\mu (A_m)_\pm\rangle,
\end{eqnarray}
   where we made use of Eq.\ (\ref{covder}).
   This derivative shifting procedure is also valid for multiple traces.
   The advantage of our new basic building blocks stems from the 
relatively simple connection between the covariant derivative 
$\nabla_\mu$ outside the block brackets and the covariant 
derivative $D_\mu$ inside 
\begin{eqnarray}
\label{nabla}
\nabla_\mu (A)_\pm&=&(D_\mu A)_\pm - \frac{1}{4}\{(D_\mu U)_-,(A)_\mp\}.
\end{eqnarray}
   Equation (\ref{nabla}) is important because 
it helps to avoid extremely tedious algebraic manipulations one had to 
perform in the old framework of Ref.\ \cite{Fearing_96}.\footnote{
A combination of shifting derivatives back and forth and interchanging 
indices of multiple derivatives is sometimes referred to as index exchange.}

Appendix A collects 23 non-redundant partial-integration relations 
(written in terms of the numbers indicated in Table II) which are used to 
eliminate dependent structures.
   Note that some of these relations might neglect contributions from 
lower levels. 
   Since we always try to find the most general Lagrangian density at each 
level, in almost all cases those lower terms do not need to be known 
explicitly.  

\section{Bianchi identities}
   As a consequence of the Jacobi identity
\begin{equation}
\label{jacobi}
[A,[B,C]]+[B,[C,A]]+[C,[A,B]]=0,
\end{equation}
certain combinations of covariant derivatives of the field strength tensors 
are not linearly independent. 
   According to Eq.~(18) of Ref.~\cite{Fearing_96}, the covariant derivative 
is defined as
\begin{equation}
\label{dfr}
D_\mu F^R_{\nu\rho}=\partial_\mu F^R_{\nu\rho}-i[R_\mu,F^R_{\nu\rho}].
\end{equation}
Now consider the linear combination
\begin{eqnarray}
D_\mu F_{\nu\rho}^R+D_\nu F^R_{\rho\mu}+ D_\rho F^R_{\mu\nu}
&\equiv&\sum_{\mbox{\scriptsize c.p.}\{\mu,\nu,\rho\}}D_\mu F^R_{\nu\rho}
=\sum_{\mbox{\scriptsize c.p.}\{\mu,\nu,\rho\}}
(\partial_\mu F^R_{\nu\rho}-i[R_\mu,F^R_{\nu\rho}])\nonumber\\
&=&\sum_{\mbox{\scriptsize c.p.}\{\mu,\nu,\rho\}}
\bigg(\partial_\mu\partial_\nu R_\rho-\partial_\mu\partial_\rho R_\nu
-i[\partial_\mu R_\nu, R_\rho]\nonumber\\
&-&i[R_\nu,\partial_\mu R_\rho]
-i[R_\mu,\partial_\nu R_\rho-\partial_\rho R_\nu]
-[R_\mu,[R_\nu, R_\rho]]\bigg)\nonumber\\
&=&0,
\label{sumcp}
\end{eqnarray}
where use of the Schwarz theorem, relabelling of indices, and the Jacobi 
identity, Eq.\ (\ref{jacobi}), has been made. 
Observe that the cyclic permutation of the indices $\mu,\nu,$ and $\rho$ has 
been denoted by $\mbox{c.p.}\{\mu,\nu,\rho\}$.
   Repeating the same arguments for the independent field strength tensor 
$F^L_{\mu\nu}$, the additional new constraints can be summarized as
\begin{equation}
\label{bi}
\sum_{\mbox{\scriptsize c.p.}\{\mu,\nu,\rho\}}D_\mu F^{L/R}_{\nu\rho}=0,
\end{equation}
which are referred to as the Bianchi identities 
(see, e.g., Refs.~\cite{Ryder,Weinberg}). 
   Equation (\ref{bi}) does not require that $F^{R/L}_{\mu\nu}$ satisfy
any equations of motion. 
   In terms of our building blocks $(D_\mu U)_-$, $(G_{\mu\nu})_+$, and 
$(H_{\mu\nu})_+$ the Bianchi identities read
\begin{eqnarray}
\label{bianchig}
\sum_{\mbox{\scriptsize c.p.}\{\mu,\nu,\rho\}}(D_\mu G_{\nu\rho})_+&=&
-\frac{1}{4}\sum_{\mbox{\scriptsize c.p.}\{\mu,\nu,\rho\}}
[(D_\mu U)_-,(H_{\nu\rho})_+],\\
\label{bianchih}
\sum_{\mbox{\scriptsize c.p.}\{\mu,\nu,\rho\}}(D_\mu H_{\nu\rho})_+&=&
-\frac{1}{4}\sum_{\mbox{\scriptsize c.p.}\{\mu,\nu,\rho\}}
[(D_\mu U)_-,(G_{\nu\rho})_+].
\end{eqnarray}   
We stress that in Ref.~\cite{Fearing_96} each term 
(though in the old notation) of the sum on the left-hand side of 
Eqs.\ (\ref{bianchig}) and (\ref{bianchih}) was treated as an 
independent element.

In Appendix D one finds 7 eliminating relations induced by Bianchi identities. 
At this point we need to stress that sometimes a structure can be eliminated
due to several arguments. Therefore, the number of non-redundant relations
belonging to a specific subprocedure can vary, while the total number of all
relations is constant. 

\section{Trace relations}
   As all the structures one can think of either involve single or multiple 
traces, one is particularly interested in finding relations between those 
traces. 
   We know from the Cayley-Hamilton theorem that any $n\times n$ matrix $A$ 
is a solution of its associated characteristical polynomial $\chi_A$. 
   For $n=2$ this statement reads
\begin{eqnarray}
\label{CalHam}
0=\chi_A (A) = A^2 - \langle A\rangle A + \det (A) 1= A^2 - \langle A\rangle A 
+ \frac{1}{2} ({\langle A\rangle}^2 - \langle A^2\rangle) 1.
\end{eqnarray}
   Setting $A=A_1+A_2$ in (\ref{CalHam}) and making use of 
$\chi_{A_1} (A_1) = 0 =\chi_{A_2} (A_2) $ one ends up with the matrix equation
\begin{eqnarray}
\label{F2}
F_2(A_1,A_2)\equiv\{A_1,A_2\} 
- \langle A_1\rangle A_2 - \langle A_2\rangle A_1 +
\langle A_1\rangle \langle A_2\rangle 1 -  \langle A_1 A_2\rangle 1=0
\end{eqnarray}
which is the central piece of information needed to derive the trace 
relations. 
   The analogous $n=3$ equation is slightly more complex 
\begin{eqnarray}
F_3(A_1,A_2,A_3)&\equiv& 
A_1\{A_2,A_3\} + A_2\{A_3,A_1\} + A_3\{A_1,A_2\}\nonumber\\
&-&\langle A_1 \rangle\{A_2,A_3\} - \langle A_2 \rangle\{A_3,A_1\} 
- \langle A_3 \rangle\{A_1,A_2\}\nonumber\\
&+&\langle A_1 \rangle\langle A_2 \rangle A_3 
+ \langle A_2 \rangle\langle A_3 \rangle A_1 
+ \langle A_3 \rangle\langle A_1 \rangle A_2\nonumber\\
&-&\langle A_1 A_2 \rangle A_3 - \langle A_3 A_1 \rangle A_2 
- \langle A_2 A_3 \rangle A_1\nonumber\\
&-&\langle A_1 A_2 A_3 \rangle 1 - \langle A_1 A_3 A_2 \rangle 1\nonumber\\ 
&+&\langle A_1 A_2 \rangle\langle A_3 \rangle 1 
+ \langle A_3 A_1 \rangle\langle A_2 \rangle 1 
+ \langle A_2 A_3 \rangle\langle A_1 \rangle 1\nonumber\\
\label{F3}
&-&\langle A_1 \rangle\langle A_2 \rangle\langle A_3 \rangle 1=0.
\end{eqnarray}
   We can now derive trace relations by simply multiplying Eq.\ (\ref{F2}) 
or Eq.\ (\ref{F3}) with another arbitrary matrix of adequate dimension and 
finally taking the trace of the whole construction, i.e., 
\begin{mathletters}
\label{Trace}
\begin{eqnarray}
\label{Tracesu2}
0&=&\langle F_2(A_1,A_2)A_3\rangle,\\
\label{tracesu3}
0&=&\langle F_3(A_1,A_2,A_3)A_4\rangle.
\end{eqnarray}
\end{mathletters}
   Note that $A_i$ may be any $n\times n$ matrix, even a string of our basic 
building blocks. 
   Although we have never come across a trace relation that could not be 
obtained in the manner explained above, we are not aware of a general proof 
showing that any kind of trace relation must be related to the Caley-Hamilton 
theorem.

   Equation (\ref{tracesu3}) is actually identical to Eq.~(A7) of 
Ref.~\cite{Fearing_96}. In Appendix E we tabulate the 19 $SU(2)$ relations 
which follow from Eq.~(\ref{Tracesu2}) plus a single $SU(3)$ relation which 
was missed in Ref.~\cite{Fearing_96}.

\section{extension of the chiral group}
   So far we have presented the most general anomalous 
$SU(N_f)_L\times SU(N_f)_R$ invariant Lagrangian density 
(respecting external fields $v_\mu, a_\mu, s,$ and $p$) 
at chiral order $p^6$ from which one can extract the $SU(2)$ or $SU(3)$ 
version by applying the respective trace relations. 
   Since $(G_{\mu\nu})_+$ and $(H_{\mu\nu})_+$ are both supposed to be 
traceless and the $SU(2)$ quark charge matrix 
$Q=\mbox{diag}(\frac{2}{3},-\frac{1}{3})$ is not, even the special case of 
electromagnetism ($L_\mu=R_\mu=-eA_\mu Q$) is not fully included in the 
general $SU(2)$ formulae. 
   As $Q$ happens to be traceless in $SU(3)$, there is no such problem in 
this case. 
   If we want to do away with this apparent weakness, we need to extend the 
chiral group to $SU(N_f)_L\times SU(N_f)_R\times U(1)_V$. 
   For $N_f=2$ the electromagnetic interaction is then fully included by 
setting 
$L_\mu=R_\mu=-\frac{e}{2}A_\mu \tau_3$ and $v^{(s)}_\mu=-\frac{e}{2}A_\mu$, 
where $v^{(s)}_\mu$ is the $U(1)_V$ gauge field. 
   Equivalently, one might argue that the traceless fields 
$L_\mu$ and $R_\mu$ have to be replaced by non-traceless ones of the
form $\tilde{L}_\mu=L_\mu+\frac{1}{3}v^{(s)}_\mu 1_{N_f\times N_f}$ and 
$\tilde{R_\mu}=R_\mu+\frac{1}{3}v^{(s)}_\mu 1_{N_f\times N_f}$, respectively. 

   The extension basically comes down to the fact that we get one more 
independent elementary building block of chiral order $p^2$, 
namely the field strength tensor 
$v^{(s)}_{\mu\nu}=\partial_\mu v^{(s)}_\nu-\partial_\nu v^{(s)}_\mu$.
   Again, we write down all (30) anomalous $p^6$ structures including at 
least one $v^{(s)}_{\mu\nu}$ (see Table~\ref{extension_list}). 
   Observe that at lower orders there is no way to construct such terms. 
   We then use the mechanisms of the previous sections to eliminate as many 
of the additional monomials as possible. 
   Since trace relations do not induce reductions in this field, 
our result is valid for any $N_f$, although we are primarily interested in 
$N_f=2$.
   While all explicit relations can be found in the Appendices, 
the final set of additional structures is shown in Table \ref{final_ex}.

\section{Comparison with our former set}
According to the present analysis the most general anomalous $SU(N_f)$
Lagrangian density of chiral order ${\cal{O}}(p^6)$, ${\cal{L}}_{6,\epsilon}$,
must not include more than 24 elements. We have started with 74 initial 
$SU(N_f)$ structures and then worked out 50 eliminating relations 
(trace relations excluded). We have tried to keep as many of the 32 
structures of Ref.~\cite{Fearing_96} (up to signs and factors) as possible.
However, some terms listed therein, such as $A_3$ and $A_{24}$, are 
unnecessarily complicated. Therefore, it turns out to be more appropriate to 
respectively replace these elements by ${\mathbf{L_{19}^{6,\epsilon}}}$ and 
${\mathbf{L_{20}^{6,\epsilon}}}$ which is justified by the total-derivative 
argument of Sec.\ III. 

The necessary reduction from 32 to 24 structures is then obtained in the 
following way.
First of all, using once more the improved partial-integration method, 
$A_{10}$ and $A_{11}$ can be shown to be redundant. 
 The application of the Bianchi identities implies the further elimination 
of $A_{1}, A_{2}, A_{9}, A_{17}, A_{18}$, and $A_{23}$. 
The exact correspondence between the new low-energy constants 
${\mathbf{L_{i}^{6,\epsilon}}}$and the old $A_j$ is shown in Table 
\ref{comparison}. 

Finally, one $SU(3)$ trace relation can be implemented to get rid of 
$A_{30}$.

\section{Summary}
   In Ref.~\cite{Fearing_96} we presented a list of 32 anomalous 
$SU(N_f)$ (the same number was valid for $SU(3)$) monomials of chiral 
order ${\cal{O}}(p^6)$ which constitute the most general mesonic ChPT 
Lagrangian 
density ${\cal{L}}_{6,\epsilon}$. 
   After carefully revisiting the entire project, our list has been found 
to be redundant. 
   Thanks to a more efficient use of partial integration, the implementation 
of so-called Bianchi identities for field strength tensors, and an additional
application of a trace relation we have ended up with 
24 $SU(N_f)$, 23 $SU(3)$, and 5 $SU(2)$ slightly modified elements. 
   Our final sets are thus considerably smaller than those proposed 
by \cite{Akhoury_91,Issler}, which are both incomplete and redundant. 
   Furthermore, we have constructed 8 additional structures which arise 
due to the extension of the chiral group to 
$SU(N_f)_L\times SU(N_f)_R\times U(1)_V$. 
   Besides the general interest, the latter group needs to be considered 
when consistently treating electromagnetic reactions in the $SU(2)$ framework.
  
\acknowledgments
   We would like to thank J.\ Bijnens and G.\ Colangelo for drawing
our attention to the Bianchi identities.
   This work was supported in part by the Deutsche Forschungsgemeinschaft 
(SFB 443) and in part by the Natural Sciences and Engineering Research Council of Canada. 
   T.~E.\ would like to thank the theory group at TRIUMF for its
hospitality and the German Academic Exchange Service (DAAD) for financial
support.

\newpage
\begin{table}[h]
\begin{center}
\begin{tabular}{|c|c|c|c|}
 our notation&  Ref.~\cite{Fearing_96}  &  Ref.~\cite{Bijnens_99}    &  Ref.~\cite{Akhoury_91}\\   
\hline  
$(\chi)_\pm$  &  $2u^\dagger [\chi]_\pm u$  &  $\chi_\pm$  &  $2M^\pm$\\
$(D_\mu U)_-$  &  $2u^\dagger [D_\mu U]_- u$  &  $-2iu_\mu$  &  $4\Delta_\mu$\\
$(G^{\mu\nu})_+$  & $2u^\dagger [G^{\mu\nu}]_+u$  &  $2f_+^{\mu\nu}$  &  $-2iF^{+\mu\nu}$\\
$(H^{\mu\nu})_+$  &  $2u^\dagger [H^{\mu\nu}]_+ u$  &  $-2f_-^{\mu\nu}$  &  $2iF^{-\mu\nu}$\\
\end{tabular}
\caption{Translation prescription.}
\label{tp}
\end{center}
\end{table}

\begin{table}[h]
\begin{center}
\begin{tabular}{|c|c|}
\# & Initial Structures\\
\hline
1 & $i\ \langle (D_\lambda D^\lambda U)_- (D_\mu U)_- (D_\nu U)_- (D_\alpha U)_- (D_\beta U)_- \rangle\epsilon^{\mu\nu\alpha\beta}$\\
2 & $i\ \langle (D_\lambda D_\mu U)_- \{ (D^\lambda U)_- (D_\nu U)_- (D_\alpha U)_- (D_\beta U)_- + (D_\beta U)_- (D_\alpha U)_- (D_\nu U)_- (D^\lambda U)_- \} \rangle\epsilon^{\mu\nu\alpha\beta}$\\
3 & $i\ \langle (D_\lambda D_\mu U)_- \{ (D_\nu U)_- (D^\lambda U)_- (D_\alpha U)_- (D_\beta U)_- + (D_\beta U)_- (D_\alpha U)_- (D^\lambda U)_- (D_\nu U)_-\} \rangle\epsilon^{\mu\nu\alpha\beta}$\\
\hline
4 & $i\ \langle (\chi)_- (D_\mu U)_- (D_\nu U)_- (D_\alpha U)_- (D_\beta U)_- \rangle\epsilon^{\mu\nu\alpha\beta}$\\
5 & $\langle (D^\lambda G_{\lambda\mu})_+ (D_\nu U)_- (D_\alpha U)_- (D_\beta U)_- \rangle\epsilon^{\mu\nu\alpha\beta}$\\
6 & $\langle (D_\lambda G_{\mu\nu})_+ \{ (D^\lambda U)_- (D_\alpha U)_- (D_\beta U)_- - (D_\beta U)_- (D_\alpha U)_- (D^\lambda U)_- \} \rangle\epsilon^{\mu\nu\alpha\beta}$\\ 
7 & $\langle (D_\lambda G_{\mu\nu})_+ (D_\alpha U)_- (D^\lambda U)_- (D_\beta U)_- \rangle\epsilon^{\mu\nu\alpha\beta}$\\
8 & $\langle (D_\mu G_{\lambda\nu})_+ \{ (D^\lambda U)_- (D_\alpha U)_- (D_\beta U)_- - (D_\beta U)_- (D_\alpha U)_- (D^\lambda U)_- \} \rangle\epsilon^{\mu\nu\alpha\beta}$\\
9 & $\langle (D_\mu G_{\lambda\nu})_+ (D_\alpha U)_- (D^\lambda U)_- (D_\beta U)_- \rangle\epsilon^{\mu\nu\alpha\beta}$\\
10 & $\langle (D_\mu G_{\nu\alpha})_+ \{ (D_\lambda U)_- (D^\lambda U)_- (D_\beta U)_- - (D_\beta U)_- (D^\lambda U)_- (D_\lambda U)_- \} \rangle\epsilon^{\mu\nu\alpha\beta}$\\ 
11 & $\langle (G_{\mu\nu})_+ \{ (D_\lambda D^\lambda U)_- (D_\alpha U)_- (D_\beta U)_- - (D_\beta U)_- (D_\alpha U)_- (D_\lambda D^\lambda U)_- \} \rangle\epsilon^{\mu\nu\alpha\beta}$\\
12 & $\langle (G_{\lambda\mu})_+ \{ (D^\lambda D_\nu U)_- (D_\alpha U)_- (D_\beta U)_- - (D_\beta U)_- (D_\alpha U)_- (D^\lambda D_\nu U)_- \} \rangle\epsilon^{\mu\nu\alpha\beta}$\\
13 & $\langle (G_{\mu\nu})_+ \{ (D_\lambda D_\alpha U)_- (D^\lambda U)_- (D_\beta U)_- - (D_\beta U)_- (D^\lambda U)_- (D_\lambda D_\alpha U)_- \} \rangle\epsilon^{\mu\nu\alpha\beta}$\\
14 & $\langle (G_{\mu\nu})_+ \{ (D_\lambda D_\alpha U)_- (D_\beta U)_- (D^\lambda U)_- - (D^\lambda U)_- (D_\beta U)_- (D_\lambda D_\alpha U)_- \} \rangle\epsilon^{\mu\nu\alpha\beta}$\\
15 & $\langle (G_{\mu\nu})_+ (D_\alpha U)_- (D_\lambda D^\lambda U)_- (D_\beta U)_- \rangle\epsilon^{\mu\nu\alpha\beta}$\\
16 & $\langle (G_{\lambda\mu})_+ (D_\nu U)_- (D^\lambda D_\alpha U)_- (D_\beta U)_- \rangle\epsilon^{\mu\nu\alpha\beta}$\\
17 & $\langle (G_{\mu\nu})_+ \{ (D_\lambda U)_- (D^\lambda D_\alpha U)_- (D_\beta U)_- - (D_\beta U)_- (D^\lambda D_\alpha U)_- (D_\lambda U)_- \} \rangle\epsilon^{\mu\nu\alpha\beta}$\\
18 & $\langle (D_\lambda D_\mu  H_{\nu\alpha})_+ \{ (D^\lambda U)_- (D_\beta U)_- + (D_\beta U)_- (D^\lambda U)_- \} \rangle\epsilon^{\mu\nu\alpha\beta}$\\
19 & $\langle (D_\lambda H_{\mu\nu})_+ \{ (D^\lambda D_\alpha U)_- (D_\beta U)_- + (D_\beta U)_- (D^\lambda D_\alpha U)_- \} \rangle\epsilon^{\mu\nu\alpha\beta}$\\
20 & $\langle (D_\mu H_{\lambda\nu})_+ \{ (D^\lambda D_\alpha U)_- (D_\beta U)_- + (D_\beta U)_- (D^\lambda D_\alpha U)_- \} \rangle\epsilon^{\mu\nu\alpha\beta}$\\
21 & $\langle (D_\mu H_{\nu\alpha})_+ \{ (D_\lambda D^\lambda U)_- (D_\beta U)_- + (D_\beta U)_- (D_\lambda D^\lambda U)_- \} \rangle\epsilon^{\mu\nu\alpha\beta}$\\
22 & $\langle (D_\mu H_{\nu\alpha})_+ \{ (D_\lambda D_\beta U)_- (D^\lambda U)_- + (D^\lambda U)_- (D_\lambda D_\beta U)_- \} \rangle\epsilon^{\mu\nu\alpha\beta}$\\
23 & $\langle (H_{\mu\nu})_+ \{ (D_\lambda D^\lambda D_\alpha U)_- (D_\beta U)_- + (D_\beta U)_- (D_\lambda D^\lambda D_\alpha U)_- \} \rangle\epsilon^{\mu\nu\alpha\beta}$\\
24 & $\langle (H_{\lambda\mu})_+ \{ (D^\lambda U)_- (D_\nu U)_- (D_\alpha U)_- (D_\beta U)_- + (D_\beta U)_- (D_\alpha U)_- (D_\nu U)_- (D^\lambda U)_- \} \rangle\epsilon^{\mu\nu\alpha\beta}$\\
25 & $\langle (H_{\lambda\mu})_+ \{ (D_\nu U)_- (D^\lambda U)_- (D_\alpha U)_- (D_\beta U)_- + (D_\beta U)_- (D_\alpha U)_- (D^\lambda U)_- (D_\nu U)_- \} \rangle\epsilon^{\mu\nu\alpha\beta}$\\
26 & $\langle (H_{\mu\nu})_+ \{ (D_\lambda U)_- (D^\lambda U)_- (D_\alpha U)_- (D_\beta U)_- + (D_\beta U)_- (D_\alpha U)_- (D^\lambda U)_- (D_\lambda U)_- \} \rangle\epsilon^{\mu\nu\alpha\beta}$\\
27 & $\langle (H_{\mu\nu})_+ \{ (D_\lambda U)_- (D_\alpha U)_- (D^\lambda U)_- (D_\beta U)_- + (D_\beta U)_- (D^\lambda U)_- (D_\alpha U)_- (D_\lambda U)_- \} \rangle\epsilon^{\mu\nu\alpha\beta}$\\
28 & $\langle (D_\lambda D_\mu H_{\nu\alpha})_+ \rangle\langle (D^\lambda U)_- (D_\beta U)_- \rangle\epsilon^{\mu\nu\alpha\beta}$\\ 
29 & $\langle (D_\lambda D^\lambda D_\mu U)_- \rangle\langle  (H_{\nu\alpha})_+ (D_\beta U)_- \rangle\epsilon^{\mu\nu\alpha\beta}$\\ 
30 & $\langle (H_{\mu\nu})_+ (D_\alpha U)_- \rangle\langle (D_\beta U)_-(D_\lambda U)_- (D^\lambda U)_- \rangle\epsilon^{\mu\nu\alpha\beta}$\\ 
31 & $\langle (H_{\mu\nu})_+ \{ (D_\lambda U)_- (D_\alpha U)_- + (D_\alpha U)_- (D_\lambda U)_- \} \rangle\langle (D^\lambda U)_-(D_\beta U)_- \rangle\epsilon^{\mu\nu\alpha\beta}$\\ 
\hline
32 & $\langle (D_\mu \chi)_+ (D_\nu H_{\alpha\beta})_+ \rangle\epsilon^{\mu\nu\alpha\beta}$\\
33 & $\langle (D_\mu \chi)_+ \{ (G_{\nu\alpha})_+ (D_\beta U)_- - (D_\beta U)_- (G_{\nu\alpha})_+ \}\rangle\epsilon^{\mu\nu\alpha\beta}$\\
34 & $\langle (D_\mu \chi)_- \{ (H_{\nu\alpha})_+ (D_\beta U)_- + (D_\beta U)_- (H_{\nu\alpha})_+ \}\rangle\epsilon^{\mu\nu\alpha\beta}$\\
\end{tabular}
\end{center}
\end{table}

\begin{table}[h]
\begin{center}
\begin{tabular}{|c|c|}
\# & continuation\\
\hline
35 & $\langle (D_\mu G_{\nu\alpha})_+ \{ (\chi)_+ (D_\beta U)_- - (D_\beta U)_- (\chi)_+ \}\rangle\epsilon^{\mu\nu\alpha\beta}$\\
36 & $\langle (D_\mu H_{\nu\alpha})_+ \{ (\chi)_- (D_\beta U)_- + (D_\beta U)_- (\chi)_- \}\rangle\epsilon^{\mu\nu\alpha\beta}$\\
37 & $i\ \langle (D^\lambda G_{\lambda\mu})_+ \{ (G_{\nu\alpha})_+ (D_\beta U)_- + (D_\beta U)_- (G_{\nu\alpha})_+ \}\rangle\epsilon^{\mu\nu\alpha\beta}$\\
38 & $i\ \langle (D^\lambda G_{\mu\nu})_+ \{ (G_{\lambda\alpha})_+ (D_\beta U)_- + (D_\beta U)_- (G_{\lambda\alpha})_+ \}\rangle\epsilon^{\mu\nu\alpha\beta}$\\
39 & $i\ \langle (D_\lambda G_{\mu\nu})_+ \{ (G_{\alpha\beta})_+ (D^\lambda U)_- + (D^\lambda U)_- (G_{\alpha\beta})_+ \}\rangle\epsilon^{\mu\nu\alpha\beta}$\\
40 & $i\ \langle (D_\mu G_{\lambda\nu})_+ \{ ({G^\lambda}_\alpha)_+ (D_\beta U)_- + (D_\beta U)_- ({G^\lambda}_\alpha)_+ \}\rangle\epsilon^{\mu\nu\alpha\beta}$\\
41 & $i\ \langle (D_\mu G_{\lambda\nu})_+ \{ (G_{\alpha\beta})_+ (D^\lambda U)_- + (D^\lambda U)_- (G_{\alpha\beta})_+ \}\rangle\epsilon^{\mu\nu\alpha\beta}$\\
42 & $i\ \langle (D_\mu G_{\nu\alpha})_+ \{ (G_{\lambda\beta})_+ (D^\lambda U)_- + (D^\lambda U)_- (G_{\lambda\beta})_+ \}\rangle\epsilon^{\mu\nu\alpha\beta}$\\
43 & $i\ \langle (D^\lambda H_{\lambda\mu})_+ \{ (H_{\nu\alpha})_+ (D_\beta U)_- + (D_\beta U)_- (H_{\nu\alpha})_+ \}\rangle\epsilon^{\mu\nu\alpha\beta}$\\
44 & $i\ \langle (D^\lambda H_{\mu\nu})_+ \{ (H_{\lambda\alpha})_+ (D_\beta U)_- + (D_\beta U)_- (H_{\lambda\alpha})_+ \}\rangle\epsilon^{\mu\nu\alpha\beta}$\\
45 & $i\ \langle (D_\lambda H_{\mu\nu})_+ \{ (H_{\alpha\beta})_+ (D^\lambda U)_- + (D^\lambda U)_- (H_{\alpha\beta})_+ \}\rangle\epsilon^{\mu\nu\alpha\beta}$\\
46 & $i\ \langle (D_\mu H_{\lambda\nu})_+ \{ ({H^\lambda}_\alpha)_+ (D_\beta U)_- + (D_\beta U)_- ({H^\lambda}_\alpha)_+ \}\rangle\epsilon^{\mu\nu\alpha\beta}$\\
47 & $i\ \langle (D_\mu H_{\lambda\nu})_+ \{ (H_{\alpha\beta})_+ (D^\lambda U)_- + (D^\lambda U)_- (H_{\alpha\beta})_+ \}\rangle\epsilon^{\mu\nu\alpha\beta}$\\
48 & $i\ \langle (D_\mu H_{\nu\alpha})_+ \{ (H_{\lambda\beta})_+ (D^\lambda U)_- + (D^\lambda U)_- (H_{\lambda\beta})_+ \}\rangle\epsilon^{\mu\nu\alpha\beta}$\\
49 & $i\ \langle (D_\lambda D^\lambda U)_- (G_{\mu\nu})_+ (G_{\alpha\beta})_+ \rangle\epsilon^{\mu\nu\alpha\beta}$\\
50 & $i\ \langle (D^\lambda D_\mu U)_- \{ (G_{\lambda\nu})_+ (G_{\alpha\beta})_+ + (G_{\alpha\beta})_+ (G_{\lambda\nu})_+ \} \rangle\epsilon^{\mu\nu\alpha\beta}$\\
51 & $i\ \langle (D_\lambda D^\lambda U)_- (H_{\mu\nu})_+ (H_{\alpha\beta})_+ \rangle\epsilon^{\mu\nu\alpha\beta}$\\
52 & $i\ \langle (D^\lambda D_\mu U)_- \{ (H_{\lambda\nu})_+ (H_{\alpha\beta})_+ + (H_{\alpha\beta})_+ (H_{\lambda\nu})_+ \} \rangle\epsilon^{\mu\nu\alpha\beta}$\\
53 & $\langle (\chi)_- \{ (G_{\mu\nu})_+ (D_\alpha U)_- (D_\beta U)_- - (D_\beta U)_- (D_\alpha U)_- (G_{\mu\nu})_+ \} \rangle\epsilon^{\mu\nu\alpha\beta}$\\
54 & $\langle (\chi)_+ \{ (H_{\mu\nu})_+ (D_\alpha U)_- (D_\beta U)_- + (D_\beta U)_- (D_\alpha U)_- (H_{\mu\nu})_+ \} \rangle\epsilon^{\mu\nu\alpha\beta}$\\
55 & $i\ \langle (G_{\lambda\mu})_+ \{ ({H^\lambda}_\nu)_+ (D_\alpha U)_- (D_\beta U)_- - (D_\beta U)_- (D_\alpha U)_- ({H^\lambda}_\nu)_+ \} \rangle\epsilon^{\mu\nu\alpha\beta}$\\
56 & $i\ \langle (G_{\lambda\mu})_+ \{ (H_{\nu\alpha})_+ (D^\lambda U)_- (D_\beta U)_- - (D_\beta U)_- (D^\lambda U)_- (H_{\nu\alpha})_+ \} \rangle\epsilon^{\mu\nu\alpha\beta}$\\
57 & $i\ \langle (G_{\lambda\mu})_+ \{ (H_{\nu\alpha})_+ (D_\beta U)_- (D^\lambda U)_- - (D^\lambda U)_- (D_\beta U)_- (H_{\nu\alpha})_+ \} \rangle\epsilon^{\mu\nu\alpha\beta}$\\
58 & $i\ \langle (G_{\mu\nu})_+ \{ (H_{\lambda\alpha})_+ (D^\lambda U)_- (D_\beta U)_- - (D_\beta U)_- (D^\lambda U)_- (H_{\lambda\alpha})_+ \} \rangle\epsilon^{\mu\nu\alpha\beta}$\\
59 & $i\ \langle (G_{\mu\nu})_+ \{ (H_{\lambda\alpha})_+ (D_\beta U)_- (D^\lambda U)_- - (D^\lambda U)_- (D_\beta U)_- (H_{\lambda\alpha})_+ \} \rangle\epsilon^{\mu\nu\alpha\beta}$\\
60 & $i\ \langle (G_{\mu\nu})_+ \{ (H_{\alpha\beta})_+ (D_\lambda U)_- (D^\lambda U)_- - (D^\lambda U)_- (D_\lambda U)_- (H_{\alpha\beta})_+ \} \rangle\epsilon^{\mu\nu\alpha\beta}$\\
61 & $\langle (\chi)_- (D_\mu U)_- (G_{\nu\alpha})_+ (D_\beta U)_-  \rangle\epsilon^{\mu\nu\alpha\beta}$\\
62 & $i\ \langle (G_{\lambda\mu})_+ \{ (D^\lambda U)_- (H_{\nu\alpha})_+ (D_\beta U)_- - (D_\beta U)_- (H_{\nu\alpha})_+ (D^\lambda U)_- \} \rangle\epsilon^{\mu\nu\alpha\beta}$\\
63 & $i\ \langle (G_{\lambda\mu})_+ (D_\nu U)_- ({H^\lambda}_\alpha)_+ (D_\beta U)_- \rangle\epsilon^{\mu\nu\alpha\beta}$\\ 
64 & $i\ \langle (G_{\mu\nu})_+ \{ (D^\lambda U)_- (H_{\lambda\alpha})_+ (D_\beta U)_- - (D_\beta U)_- (H_{\lambda\alpha})_+ (D^\lambda U)_- \} \rangle\epsilon^{\mu\nu\alpha\beta}$\\
65 & $\langle (D_\mu \chi)_- \rangle\langle (H_{\nu\alpha})_+ (D_\beta U)_- \rangle\epsilon^{\mu\nu\alpha\beta}$\\ 
66 & $\langle (\chi)_- \rangle\langle (D_\mu H_{\nu\alpha})_+ (D_\beta U)_-  \rangle\epsilon^{\mu\nu\alpha\beta}$\\ 
67 & $\langle (\chi)_- \rangle\langle (G_{\mu\nu})_+ (D_\alpha U)_- (D_\beta U)_- \rangle\epsilon^{\mu\nu\alpha\beta}$\\
68 & $\langle (\chi)_+ (D_\mu U)_- \rangle\langle (H_{\nu\alpha})_+ (D_\beta U)_- \rangle\epsilon^{\mu\nu\alpha\beta}$\\ 
\hline
69 & $i\ \langle (\chi)_+ \{ (G_{\mu\nu})_+ (H_{\alpha\beta})_+ - (H_{\alpha\beta})_+ (G_{\mu\nu})_+ \} \rangle\epsilon^{\mu\nu\alpha\beta}$\\
70 & $i\ \langle (\chi)_- (G_{\mu\nu})_+ (G_{\alpha\beta})_+ \rangle\epsilon^{\mu\nu\alpha\beta}$\\
71 & $i\ \langle (\chi)_- (H_{\mu\nu})_+ (H_{\alpha\beta})_+ \rangle\epsilon^{\mu\nu\alpha\beta}$\\
72 & $\langle (G_{\mu\nu})_+ \{ (G_{\lambda\alpha})_+ ({H^\lambda}_\beta)_+ + {(H^\lambda}_\beta)_+ (G_{\lambda\alpha})_+ \} \rangle\epsilon^{\mu\nu\alpha\beta}$\\
73 & $i\ \langle (\chi)_- \rangle\langle (G_{\mu\nu})_+ (G_{\alpha\beta})_+ \rangle\epsilon^{\mu\nu\alpha\beta}$\\ 
74 & $i\ \langle (\chi)_- \rangle\langle (H_{\mu\nu})_+ (H_{\alpha\beta})_+ \rangle\epsilon^{\mu\nu\alpha\beta}$\\
\end{tabular}
\caption{List of 74 anomalous $SU(N_f)$ structures of order ${\cal{O}}(p^6)$ we start with. Besides the hierarchy argument discussed in the text (the four different categories are separated by simple lines) our ordering scheme is supposed to respect the rule: single traces $>$ multiple traces.}
\label{structure_list}
\end{center}
\end{table}

\begin{table}[h]
\begin{center}
\begin{tabular}{|c|c|}
\# & Initial Structures\\
\hline
$1'$ & $\partial^\lambda v^{(s)}_{\lambda\mu}\langle (D_\nu U)_- (D_\alpha U)_- (D_\beta U)_- \rangle\epsilon^{\mu\nu\alpha\beta}$\\
$2'$ & $\partial_\lambda v^{(s)}_{\mu\nu}\langle (D^\lambda U)_- (D_\alpha U)_- (D_\beta U)_- \rangle\epsilon^{\mu\nu\alpha\beta}$\\
$3'$ & $\partial_\mu v^{(s)}_{\lambda\nu}\langle (D^\lambda U)_- (D_\alpha U)_- (D_\beta U)_- \rangle\epsilon^{\mu\nu\alpha\beta}$\\
$4'$ & $v^{(s)}_{\lambda\mu}\langle (D^\lambda D_\nu U)_- (D_\alpha U)_- (D_\beta U)_- \rangle\epsilon^{\mu\nu\alpha\beta}$\\
$5'$ & $v^{(s)}_{\mu\nu}\langle (D_\lambda D^\lambda U)_- (D_\alpha U)_- (D_\beta U)_- \rangle\epsilon^{\mu\nu\alpha\beta}$\\
$6'$ & $v^{(s)}_{\mu\nu}\langle (D_\lambda D_\alpha U)_- \{ (D^\lambda U)_- (D_\beta U)_- - (D_\beta U)_- (D^\lambda U)_- \} \rangle\epsilon^{\mu\nu\alpha\beta}$\\
\hline
$7'$ & $i\ \partial^\lambda v^{(s)}_{\lambda\mu}\langle (G_{\nu\alpha})_+ (D_\beta U)_- \rangle\epsilon^{\mu\nu\alpha\beta}$\\
$8'$ & $i\ \partial^\lambda v^{(s)}_{\mu\nu}\langle (G_{\lambda\alpha})_+ (D_\beta U)_- \rangle\epsilon^{\mu\nu\alpha\beta}$\\
$9'$ & $i\ \partial_\lambda v^{(s)}_{\mu\nu}\langle (G_{\alpha\beta})_+ (D^\lambda U)_- \rangle\epsilon^{\mu\nu\alpha\beta}$\\
$10'$ & $i\ \partial_\mu v^{(s)}_{\lambda\nu}\langle ({G^\lambda}_\alpha)_+ (D_\beta U)_- \rangle\epsilon^{\mu\nu\alpha\beta}$\\
$11'$ & $i\ \partial_\mu v^{(s)}_{\lambda\nu}\langle (G_{\alpha\beta})_+ (D^\lambda U)_- \rangle\epsilon^{\mu\nu\alpha\beta}$\\
$12'$ & $i\ \partial_\mu v^{(s)}_{\nu\alpha}\langle (G_{\lambda\beta})_+ (D^\lambda U)_- \rangle\epsilon^{\mu\nu\alpha\beta}$\\
$13'$ & $i\ v^{(s)}_{\mu\nu}\langle (D^\lambda G_{\lambda\alpha})_+ (D_\beta U)_- \rangle\epsilon^{\mu\nu\alpha\beta}$\\
$14'$ & $i\ v^{(s)}_{\lambda\mu}\langle (D^\lambda G_{\nu\alpha})_+ (D_\beta U)_- \rangle\epsilon^{\mu\nu\alpha\beta}$\\
$15'$ & $i\ v^{(s)}_{\mu\nu}\langle (D_\lambda G_{\alpha\beta})_+ (D^\lambda U)_- \rangle\epsilon^{\mu\nu\alpha\beta}$\\
$16'$ & $i\ v^{(s)}_{\lambda\mu}\langle (D_\nu {G^\lambda}_\alpha)_+ (D_\beta U)_- \rangle\epsilon^{\mu\nu\alpha\beta}$\\
$17'$ & $i\ v^{(s)}_{\mu\nu}\langle (D_\alpha G_{\lambda\beta})_+ (D^\lambda U)_- \rangle\epsilon^{\mu\nu\alpha\beta}$\\
$18'$ & $i\ v^{(s)}_{\lambda\mu}\langle (D_\nu G_{\alpha\beta})_+ (D^\lambda U)_- \rangle\epsilon^{\mu\nu\alpha\beta}$\\
$19'$ & $i\ v^{(s)}_{\mu\nu}\langle (G_{\alpha\beta})_+ (D_\lambda D^\lambda U)_- \rangle\epsilon^{\mu\nu\alpha\beta}$\\
$20'$ & $i\ v^{(s)}_{\lambda\mu}\langle (G_{\nu\alpha})_+ (D^\lambda D_\beta U)_- \rangle\epsilon^{\mu\nu\alpha\beta}$\\
$21'$ & $i\ v^{(s)}_{\mu\nu}\langle (G_{\lambda\alpha})_+ (D^\lambda D_\beta U)_- \rangle\epsilon^{\mu\nu\alpha\beta}$\\
$22'$ & $i\ v^{(s)}_{\lambda\mu}\langle ({H^\lambda}_\nu)_+ (D_\alpha U)_- (D_\beta U)_- \rangle\epsilon^{\mu\nu\alpha\beta}$\\ 
$23'$ & $i\ v^{(s)}_{\lambda\mu}\langle (H_{\nu\alpha})_+ \{ (D^\lambda U)_- (D_\beta U)_- - (D_\beta U)_- (D^\lambda U)_- \} \rangle\epsilon^{\mu\nu\alpha\beta}$\\ 
$24'$ & $i\ v^{(s)}_{\mu\nu}\langle (H_{\lambda\alpha})_+ \{ (D^\lambda U)_- (D_\beta U)_- - (D_\beta U)_- (D^\lambda U)_- \} \rangle\epsilon^{\mu\nu\alpha\beta}$\\ 
$25'$ & $v^{(s)}_{\mu\nu}\langle (\chi)_- (D_\alpha U)_- (D_\beta U)_- \rangle\epsilon^{\mu\nu\alpha\beta}$\\
\hline
$26'$ & $i\ v^{(s)}_{\mu\nu}\langle (\chi)_- (G_{\alpha\beta})_+ \rangle\epsilon^{\mu\nu\alpha\beta}$\\
$27'$ & $v^{(s)}_{\lambda\mu}\langle ({G^\lambda}_\nu)_+ (H_{\alpha\beta})_+ \rangle\epsilon^{\mu\nu\alpha\beta}$\\
$28'$ & $v^{(s)}_{\lambda\mu}\langle (G_{\nu\alpha})_+ ({H^\lambda}_\beta)_+ \rangle\epsilon^{\mu\nu\alpha\beta}$\\
$29'$ & $v^{(s)}_{\mu\nu}\langle (G_{\lambda\alpha})_+ ({H^\lambda}_\beta)_+ \rangle\epsilon^{\mu\nu\alpha\beta}$\\
$30'$ & $i\ v^{(s)}_{\mu\nu} v^{(s)}_{\alpha\beta} \langle (\chi)_-\rangle\epsilon^{\mu\nu\alpha\beta}$\\
\end{tabular}
\caption{List of 30 additional anomalous $p^6$ structures due to the extension of the chiral group.}
\label{extension_list}
\end{center}
\end{table}

\newpage
\begin{table}[h]
\begin{center}
\begin{tabular}{|c|c|c|c|c|}
${\mathbf{L_{i}^{6,\epsilon}}}$ & $\#$ & $SU(N_f)$ & $SU(3)$ & $SU(2)$ \\
\hline
3 & 69 & $i\ \langle (\chi)_+ \{ (G_{\mu\nu})_+ (H_{\alpha\beta})_+ -  \mbox{rev} \} \rangle\epsilon^{\mu\nu\alpha\beta}$ & $\times$ & $\times$ \\
8 & 70 & $i\ \langle (\chi)_- (G_{\mu\nu})_+ (G_{\alpha\beta})_+ \rangle\epsilon^{\mu\nu\alpha\beta}$ & $\times$ & $\times$ \\
9 & 73 & $i\ \langle (\chi)_- \rangle\langle (G_{\mu\nu})_+ (G_{\alpha\beta})_+ \rangle\epsilon^{\mu\nu\alpha\beta}$ & $\times$ & ----- \\ 
19 & 37 & $i\ \langle (D^\lambda G_{\lambda\mu})_+ \{ (G_{\nu\alpha})_+ (D_\beta U)_- + \mbox{rev} \}\rangle\epsilon^{\mu\nu\alpha\beta}$ & $\times$ & ----- \\
\hline
1 & 54 & $\langle (\chi)_+ \{ (H_{\mu\nu})_+ (D_\alpha U)_- (D_\beta U)_- + \mbox{rev} \} \rangle\epsilon^{\mu\nu\alpha\beta}$ & $\times$ & $\times$ \\
2 & 68 & $\langle (\chi)_+ (D_\mu U)_- \rangle\langle (D_\nu U)_- (H_{\alpha\beta})_+ \rangle\epsilon^{\mu\nu\alpha\beta}$ & $\times$ & ----- \\ 
5 & 53 & $\langle (\chi)_- \{ (G_{\mu\nu})_+ (D_\alpha U)_- (D_\beta U)_- -  \mbox{rev} \} \rangle\epsilon^{\mu\nu\alpha\beta}$ & $\times$ & ----- \\
6 & 61 & $\langle (\chi)_- (D_\mu U)_- (G_{\nu\alpha})_+ (D_\beta U)_-  \rangle\epsilon^{\mu\nu\alpha\beta}$ & $\times$ & $\times$ \\
7 & 67 &\ $\langle (\chi)_- \rangle\langle (G_{\mu\nu})_+ (D_\alpha U)_- (D_\beta U)_- \rangle\epsilon^{\mu\nu\alpha\beta}$ & $\times$ & ----- \\
13 & $14^s$ & $\langle (G_{\mu\nu})_+ \{ (D^\lambda D_\alpha U)^s_- (D_\beta U)_- (D_\lambda U)_- - \mbox{rev} \} \rangle\epsilon^{\mu\nu\alpha\beta}$ & $\times$ & ----- \\
14 & $13^s$ & $\langle (G_{\mu\nu})_+ \{ (D_\lambda D_\alpha U)^s_- (D^\lambda U)_- (D_\beta U)_- - \mbox{rev} \} \rangle\epsilon^{\mu\nu\alpha\beta}$ & $\times$ & ----- \\
\hline
10 & 71 & $i\ \langle (\chi)_- (H_{\mu\nu})_+ (H_{\alpha\beta})_+ \rangle\epsilon^{\mu\nu\alpha\beta}$ & $\times$ & $\times$ \\ 
11 & 74 & $i\ \langle (\chi)_- \rangle\langle (H_{\mu\nu})_+ (H_{\alpha\beta})_+ \rangle\epsilon^{\mu\nu\alpha\beta}$ & $\times$ & ----- \\ 
20 & 43 & $i\ \langle (D^\lambda H_{\lambda\mu})_+ \{ (H_{\nu\alpha})_+ (D_\beta U)_- + \mbox{rev} \}\rangle\epsilon^{\mu\nu\alpha\beta}$ & $\times$ & ----- \\ 
21 & 55 & $i\ \langle (G_{\lambda\mu})_+ \{ ({H^\lambda}_\nu)_+ (D_\alpha U)_- (D_\beta U)_- - \mbox{rev} \}\rangle\epsilon^{\mu\nu\alpha\beta}$ & $\times$ & ----- \\
22 & 57 & $i\ \langle (G_{\lambda\mu})_+ \{ (H_{\nu\alpha})_+ (D_\beta U)_- (D^\lambda U)_- - \mbox{rev} \}\rangle\epsilon^{\mu\nu\alpha\beta}$ & $\times$ & ----- \\
23 & 60 & $i\ \langle (G_{\mu\nu})_+ \{ (H_{\alpha\beta})_+ (D_\lambda U)_- (D^\lambda U)_- - \mbox{rev} \}\rangle\epsilon^{\mu\nu\alpha\beta}$ & $\times$ & ----- \\
24 & 63 & $i\ \langle (G_{\lambda\mu})_+ (D_\nu U)_- ({H^\lambda}_\alpha)_+ (D_\beta U)_- \rangle\epsilon^{\mu\nu\alpha\beta}$ & $\times$ & ----- \\ 
\hline
4 & 4 & $i\ \langle (\chi)_- (D_\mu U)_- (D_\nu U)_- (D_\alpha U)_- (D_\beta U)_- \rangle\epsilon^{\mu\nu\alpha\beta}$ & $\times$ & ----- \\ 
12 & $2^s$ & $i\ \langle (D_\lambda D_\mu U)^s_- \{ (D^\lambda U)_- (D_\nu U)_- (D_\alpha U)_- (D_\beta U)_- + \mbox{rev} \} \rangle\epsilon^{\mu\nu\alpha\beta}$ & $\times$ & ----- \\
\hline
15 & 27 & $\langle (H_{\mu\nu})_+ \{ (D_\lambda U)_- (D_\alpha U)_- (D^\lambda U)_- (D_\beta U)_- + \mbox{rev} \} \rangle\epsilon^{\mu\nu\alpha\beta}$ & ----- & ----- \\
16 & 26 & $\langle (H_{\mu\nu})_+ \{ (D_\lambda U)_- (D^\lambda U)_- (D_\alpha U)_- (D_\beta U)_- + \mbox{rev} \} \rangle\epsilon^{\mu\nu\alpha\beta}$ & $\times$ & -----  \\
17 & 31 & $\langle (H_{\mu\nu})_+ \{ (D_\lambda U)_- (D_\alpha U)_- + \mbox{rev} \} \rangle\langle (D^\lambda U)_-(D_\beta U)_- \rangle\epsilon^{\mu\nu\alpha\beta}$ & $\times$ & ----- \\ 
18 & 30 & $\langle (H_{\mu\nu})_+ (D_\alpha U)_- \rangle\langle (D_\beta U)_-(D_\lambda U)_- (D^\lambda U)_- \rangle\epsilon^{\mu\nu\alpha\beta}$ & $\times$ & ----- \\ 
\end{tabular}
\caption{The final $SU(N_f)$ set. The remaining structures and their associated low-energy constants (LECs) have been relabelled: terms with a $(\chi)_\pm$ block from ${\mathbf{L_{1}^{6,\epsilon}}}$ to ${\mathbf{L_{11}^{6,\epsilon}}}$ and those without from ${\mathbf{L_{12}^{6,\epsilon}}}$ to ${\mathbf{L_{24}^{6,\epsilon}}}$. Afterwards they have been classified according to their leading-order expansion in terms of Goldstone boson and electromagnetic fields, see Table~\ref{classi}. The abbreviation ``rev'' stands for the reverse order.
Trace relations lead to the corresponding $SU(3)$ and $SU(2)$ sets, which are indicated by a $\times$ in the respective columns.
Recall that under the traces the new $(\cdots)_\pm$ blocks are equivalent to the old $[\cdots]_\pm$ up to a factor $2$.}
\label{final_set}
\end{center}
\end{table}

\begin{table}[h]
\begin{center}
\begin{tabular}{|c|c|}
reaction type & LECs \\  
\hline  
$\phi+2\gamma^\ast$   & $3, 8, 9, 19$ \\
$3\phi+\gamma^\ast$   & $1, 2, 5, 6, 7, 13, 14$ \\
$3\phi+2\gamma^\ast$  & $ 10, 11, 20, 21, 22, 23, 24$ \\
$5\phi$          & $4, 12$ \\
$5\phi+\gamma^\ast$   & $15, 16, 17, 18$ \\
\end{tabular}
\caption{Classification of LECs according to their leading-order contribution allowing at most electromagnetic external fields; the photons may be either virtual or real. This classification defines the ordering scheme in Table \ref{final_set}.}
\label{classi}
\end{center}
\end{table}

\begin{table}[h]
\begin{center}
\begin{tabular}{|c|c|c|}
${\mathbf{L_{i}^{6,\epsilon}}}$ & \# & additional structures\\
\hline
$2'$ & $26'$ & $i\ v^{(s)}_{\mu\nu}\langle (\chi)_- (G_{\alpha\beta})_+ \rangle\epsilon^{\mu\nu\alpha\beta}$\\
$3'$ & $30'$ & $i\ v^{(s)}_{\mu\nu} v^{(s)}_{\alpha\beta} \langle (\chi)_-\rangle\epsilon^{\mu\nu\alpha\beta}$\\
$5'$ & $13'$ & $i\ v^{(s)}_{\mu\nu}\langle (D^\lambda G_{\lambda\alpha})_+ (D_\beta U)_- \rangle\epsilon^{\mu\nu\alpha\beta}$\\
$6'$ & $7'$ & $i\ \partial^\lambda v^{(s)}_{\lambda\mu}\langle (G_{\nu\alpha})_+ (D_\beta U)_- \rangle\epsilon^{\mu\nu\alpha\beta}$\\
\hline
$1'$ & $25'$ & $v^{(s)}_{\mu\nu}\langle (\chi)_- (D_\alpha U)_- (D_\beta U)_- \rangle\epsilon^{\mu\nu\alpha\beta}$\\ 
$4'$ & $1'$ & $\partial^\lambda v^{(s)}_{\lambda\mu}\langle (D_\nu U)_- (D_\alpha U)_- (D_\beta U)_- \rangle\epsilon^{\mu\nu\alpha\beta}$\\
\hline
$7'$ & $22'$ & $i\ v^{(s)}_{\lambda\mu}\langle ({H^\lambda}_\nu)_+ (D_\alpha U)_- (D_\beta U)_- \rangle\epsilon^{\mu\nu\alpha\beta}$\\
\hline
$8'$ & $29'$ & $v^{(s)}_{\mu\nu}\langle (G_{\lambda\alpha})_+ ({H^\lambda}_\beta)_+ \rangle\epsilon^{\mu\nu\alpha\beta}$\\
\end{tabular}
\caption{Final set of additional structures valid for arbitrary $N_f$.}
\label{final_ex}
\end{center}
\end{table}

\newpage
\begin{table}[h]
\begin{center}
\begin{tabular}{|c|c|c|c|c|c|c|c|c|}
 LECs & ${\mathbf{L_{1}^{6,\epsilon}}}$ & ${\mathbf{L_{2}^{6,\epsilon}}}$ & ${\mathbf{L_{3}^{6,\epsilon}}}$ & ${\mathbf{L_{4}^{6,\epsilon}}}$ & ${\mathbf{L_{5}^{6,\epsilon}}}$ & ${\mathbf{L_{6}^{6,\epsilon}}}$ & ${\mathbf{L_{7}^{6,\epsilon}}}$ & ${\mathbf{L_{8}^{6,\epsilon}}}$ \\
\hline
Ref.~\cite{Fearing_96} & $16 A_{15}$ & $-16 A_{16}$ & $-8 A_5$ & $16 A_{28}$ & $16 A_{12}$ & $-16 A_{13}$ & $16 A_{14}$ & $8 A_{4}$ \\
\hline\hline
LECs & ${\mathbf{L_{9}^{6,\epsilon}}}$ & ${\mathbf{L_{10}^{6,\epsilon}}}$ & ${\mathbf{L_{11}^{6,\epsilon}}}$ & ${\mathbf{L_{12}^{6,\epsilon}}}$ & ${\mathbf{L_{13}^{6,\epsilon}}}$ & ${\mathbf{L_{14}^{6,\epsilon}}}$ & ${\mathbf{L_{15}^{6,\epsilon}}}$ & ${\mathbf{L_{16}^{6,\epsilon}}}$ \\
\hline
Ref.~\cite{Fearing_96} & $8 A_6$ & $8 A_{25}$ & $8 A_{26}$ & $32 A_{27}$ & $16 A_8$ & $16 A_7$ & $-32 A_{30}$ & $-32 A_{29}$ \\
\hline\hline
LECs & ${\mathbf{L_{17}^{6,\epsilon}}}$ & ${\mathbf{L_{18}^{6,\epsilon}}}$ & ${\mathbf{L_{19}^{6,\epsilon}}}$ & ${\mathbf{L_{20}^{6,\epsilon}}}$ & ${\mathbf{L_{21}^{6,\epsilon}}}$ & ${\mathbf{L_{22}^{6,\epsilon}}}$ & ${\mathbf{L_{23}^{6,\epsilon}}}$ & ${\mathbf{L_{24}^{6,\epsilon}}}$ \\
\hline
Ref.~\cite{Fearing_96} & $32 A_{32}$ & $32 A_{31}$ & n.d.c. & n.d.c. & $16 A_{20}$ & $-16 A_{22}$ & $16 A_{19}$ & $16 A_{21}$ \\
\end{tabular}
\caption{Numerical correspondence between the $SU(N_f)$ low-energy constants (LECs) defined in Table \ref{final_set} and those of our former set. Observe that ``n.d.c'' stands for ``no direct correspondence.''}
\label{comparison}
\end{center}
\end{table}

\appendix
\section{Relations induced by partial integration}
\begin{eqnarray}
(1)- (2) + (3)&=&0\nonumber,\\ 
(5) + (12) + (16)&=&0\nonumber,\\
(6) + (11) + (14)+(17)&=&0\nonumber,\\
(7) + (13) + (15)&=&0\nonumber,\\
(8) - (12)&=&0\nonumber,\\
(9) + (16)&=&0\nonumber,\\
(10) + (13) + (17)&=&0\nonumber,\\ 
(18) + (21) + (22) +\frac{1}{16}(26) +\frac{1}{8}(27)&=&0\nonumber,\\
(20) - \frac{1}{16}(24) - \frac{1}{16}(25)&=&0\nonumber,\\
(22) + \frac{1}{16}(26)&=&0\nonumber,\\
(19) + (23) - \frac{1}{16}(26) - \frac{1}{8}(27)&=&0\nonumber,\\
(21) + (23) - \frac{1}{8}(26)&=&0\nonumber,\\ 
(28) - \frac{1}{8}(31)&=&0\nonumber,\\
(29) + \frac{1}{4}(30)&=&0\nonumber,\\ 
(32) + \frac{1}{4}(36) + \frac{1}{16}(54)&=&0\nonumber,\\
(32) - \frac{1}{4}(34)&=&0\nonumber,\\
(37) + (38) - (50)&=&0\nonumber,\\
(39) + (49)&=&0\nonumber,\\
(41) + (42) + (50)&=&0\nonumber,\\
(43) + (44) - (52)&=&0\nonumber,\\
(45) + (51)&=&0\nonumber,\\
(47) + (48) + (52)&=&0\nonumber,\\ 
(65) + (66) - \frac{1}{2}(68)&=&0;\\ \nonumber\\
(1') + 3(4')&=&0\nonumber,\\ 
(2') + (5') - (6')&=&0\nonumber,\\ 
(7') + (14') + (20')&=&0\nonumber,\\ 
(8') +  (13') + (21')&=&0\nonumber,\\ 
(9') + (15') + (19')&=&0\nonumber,\\ 
(10') - (16')&=&0. 
\end{eqnarray}
Note that some of these relations neglect contributions from 
lower levels in the hierarchy.

\section{Relations induced by EOM}
\begin{eqnarray}
(1)&=&(4)\nonumber,\\
(11)&=&(53) - \frac{2}{N_f}(67)\nonumber,\\
(15)&=&-(61) - \frac{1}{N_f}(67)\nonumber,\\
(21)&=&(36) - \frac{2}{N_f}(66)\nonumber,\\
(49)&=&(70) - \frac{1}{N_f}(73)\nonumber,\\
(51)&=&(71) - \frac{1}{N_f}(74);\\ \nonumber\\
(5')&=&(25')\nonumber,\\
(19')&=&(26').
\end{eqnarray}

\section{Epsilon relations}
\begin{eqnarray}
(11) + 2(12) - (13) + (14)&=&0\nonumber,\\
(15) - 2(16) - (17)&=&0\nonumber,\\
(24) - (25)&=&0\nonumber,\\
2(24) + (26) - (27)&=&0\nonumber,\\
(38) - 2(40) + (42)&=&0\nonumber,\\
(44) - 2(46) + (48)&=&0\nonumber,\\
(49) - 2(50)&=&0\nonumber,\\
(51) - 2(52)&=&0\nonumber,\\
2(55) + (56) - (57)&=&0\nonumber,\\
2(55) - (58) + (59)&=&0\nonumber,\\
2(56) + 2(58) + (60)&=&0\nonumber,\\ 
(62) - 2(63)&=&0\nonumber,\\
(62) + (64)&=&0\nonumber,\\
(72)&=&0;\\ \nonumber\\
2(1') + 3(2')&=&0\nonumber,\\
(1') + 3(3')&=&0\nonumber,\\
2(7') + 2(8') + (9')&=&0\nonumber,\\
2(13') + 2(14') + (15')&=&0\nonumber,\\
(13') + 2(16') + (17')&=&0\nonumber,\\
(14') - 2(16') - (18')&=&0\nonumber,\\
2(22') + (23')&=&0\nonumber,\\
2(22') - (24')&=&0\nonumber,\\
(27') - (29')&=&0\nonumber,\\
(28') + (29')&=&0.
\end{eqnarray}

\section{Relations induced by Bianchi identities}
\begin{eqnarray}
(10) - \frac{1}{4}(26)&=&0\nonumber,\\ 
(32) - \frac{1}{4}(33)&=&0\nonumber,\\ 
(35) + \frac{1}{4}(54)&=&0\nonumber,\\ 
(36) - \frac{1}{4}(53) + \frac{1}{2}(61)&=&0\nonumber,\\ 
(42) + \frac{1}{4}(57) + \frac{1}{4}(62)&=&0\nonumber,\\
(48) - \frac{1}{4}(58) - \frac{1}{4}(64)&=&0\nonumber,\\ 
(66) - \frac{1}{2}(67)&=&0;\\ \nonumber\\
(8') - 2(10')&=&0\nonumber,\\
(9') - 2(11')&=&0\nonumber,\\
(12')&=&0\nonumber,\\
(18') + \frac{1}{4}(23')&=&0.
\end{eqnarray}

\section{Trace relations}
$SU(2)$:
\begin{eqnarray}
(2)&=&0\nonumber,\\
(4)&=&0\nonumber,\\
(12)&=&0\nonumber,\\
(13)&=&0\nonumber,\\
(24)&=&0\nonumber,\\
(26)&=&0\nonumber,\\
(30)&=&0\nonumber,\\
(31)&=&0\nonumber,\\
(37)&=&0\nonumber,\\
(43)&=&0\nonumber,\\
(53)+2(61)&=&0\nonumber,\\
(53)-(67)&=&0\nonumber,\\
(54)+2(68)&=&0\nonumber,\\
(58)&=&0\nonumber,\\
(59)&=&0\nonumber,\\
(60)&=&0\nonumber,\\
(63)&=&0\nonumber,\\
2(70)-(73)&=&0\nonumber,\\
2(71)-(74)&=&0;
\end{eqnarray}
$SU(3)$:
\begin{eqnarray}
(26)+(27)+2(30)+(31)=0.
\end{eqnarray}


\frenchspacing
\begin{thebibliography}{References}
\bibitem{Weinberg_79} 
   S. Weinberg, Physica {\bf 96A}, 327 (1979).
\bibitem{Gasser_84} 
   J. Gasser and H. Leutwyler, Ann. Phys. {\bf 158}, 142 (1984).
\bibitem{Gasser_85} 
   J. Gasser and H. Leutwyler, Nucl. Phys. {\bf B250}, 465 (1985).
\bibitem{Bernstein_98}
   A. M. Bernstein, D. Drechsel, and Th. Walcher (Eds.),
   {\em Chiral Dynamics: Theory and Experiment}, Proceedings, Mainz,
   Germany, 1997 (Springer, Berlin, 1998). 
\bibitem{Nambu_60} 
   Y. Nambu, Phys. Rev. Lett. {\bf 4}, 380 (1960).
\bibitem{Goldstone_61}
   J. Goldstone, Nuovo Cim. {\bf 19}, 154 (1961).
\bibitem{Goldstone_62}
   J. Goldstone, A. Salam, and S. Weinberg, Phys. Rev. {\bf 127},  965 (1962).
\bibitem{Stern_98}
   J. Stern, {\em Light Quark Masses and Condensates in QCD}, in Ref.
\cite{Bernstein_98}.
\bibitem{Fearing_96}
   H.\ W.\ Fearing and S.\ Scherer, Phys.\ Rev.\ D {\bf 53}, 315 (1996).
\bibitem{Bijnens_99}  
   J. Bijnens, G. Colangelo, and G. Ecker, JHEP {\bf 9902}, 20 (1999).
\bibitem{Akhoury_91}  
   R. Akhoury and A. Alfakih, Ann. Phys. (N.Y.) {\bf 210}, 81 (1991).
\bibitem{Ryder}
   L.\ H.\ Ryder, {\em Quantum Field Theory} (Cambridge University Press,
Cambridge, England, 1986), Chap.\ 3.6.
\bibitem{Weinberg}
   S.\ Weinberg, {\em The Quantum Theory of Fields, Vol.\ II}
(Cambridge University Press, Cambridge, England, 1996), Chap.\ 15.3.
\bibitem{Issler}
   D.\ Issler, preprint SLAC-PUB-4943, 1990 (unpublished).
\end{thebibliography}
\end{document}